\documentclass[reprint,aps,pra]{revtex4-1}
\usepackage{graphicx}

\begin{document}

\newcommand{\micron}{\ensuremath{\mu\mathrm{m}}}
\newcommand{\Ohm}{\ensuremath{\Omega}}
\newcommand{\Gn}{\ensuremath{G_0}}
\newcommand{\Zn}{\ensuremath{Z_0}}
\newcommand{\Ld}{\ensuremath{d}}
\newcommand{\Ll}{\ensuremath{l}}
\newcommand{\loss}{\ensuremath{\alpha}}
\newcommand{\pvelo}{\ensuremath{\beta}}
\newcommand{\Zd}{\ensuremath{Z_d}}
\newcommand{\Zload}{\ensuremath{Z_{\rm{Load}}}}
\newcommand{\reflcabs}{\ensuremath{|S_{11}|}}
\newcommand{\reflc}{\ensuremath{S_{11}}}
\newcommand{\fres}{\ensuremath{\nu_{\rm{r}}}}
\newcommand{\rbj}{\ensuremath{R_{\rm{BJ}}}}
\newcommand{\zbj}{\ensuremath{Z_{\rm{BJ}}}}
\newcommand{\qwl}{\ensuremath{\lambda/4}}

\author{Gabriel Puebla-Hellmann}
\email{gabriepu@phys.ethz.ch}
\author{Andreas Wallraff}

\date{\today}
\affiliation{Department of Physics, ETH Zurich, 8093 Zurich, Switzerland}

\title{Realization of GHz-frequency impedance matching circuits for nano-scale devices}

\begin{abstract}
Integrating nano-scale objects, such as single molecules or carbon nanotubes, into impedance transformers and performing radio-frequency measurements allows for high time-resolution transport measurements with improved signal-to-noise ratios. The realization of such transformers implemented with superconducting transmission lines for the 2-10 GHz frequency range is presented here. Controlled electromigration of an integrated gold break junction is used to characterize a 6 GHz impedance matching device. The real part of the RF impedance of the break junction extracted from microwave reflectometry at a maximum bandwidth of 45 MHz of the matching circuit is in good agreement with the measured direct current resistance.
\end{abstract}

\maketitle

Research into three-terminal single molecule devices has made substantial progress in the past decade.  Not only their feasibility was demonstrated but also phenomena such as the Kondo Effect \cite{Park02,Liang02,Roch08}, single spin effects \cite{Grose08} and orbital gating \cite{Song09} were observed. The large resistances of such devices (20 k\Ohm\ to G\Ohm) together with the stray capacitances of the wiring create a low-pass filter for measurement signals with a bandwidth in the kHz range, placing an upper bound on the attainable time resolution. This limitation can be overcome by integrating the device into a radio-frequency (RF) resonant circuit, transforming the device impedance to 50 \Ohm\ and measuring the reflected electromagnetic wave.
Increasing the bandwidth of transport measurements in nano-scale objects not only allows reducing the time an individual measurement requires, but also allows the investigation of transport dynamics using time-resolved measurements.
Pioneered in the RF single electron transistor \cite{Schoelkopf98}, this method has also been applied to time-resolved electron counting and charge sensing using both quantum point contacts \cite{Qin06,Cassidy07,Reilly07} and quantum dots \cite{Barthel10}, as well as displacement detection using an atomic point contact \cite{Flowers-Jacobs07}.

The simplest resonant circuit for such an application is the $L$-section \cite{Pozar05} of two reactive components, for example a series inductor shunted by a capacitor, which can be implemented either with lumped or distributed elements.
While lumped element implementations
are usually easy to realize at frequencies up to several 100's of MHz, or even in the 1-2 GHz range using on-chip superconducting inductors \cite{Xue2007,Fong2012}, they suffer from parasitic capacitances, limiting both the maximum operating frequency and the maximum load which can be transformed to 50 \Ohm.
Distributed element resonant circuits, implemented in, for example, Rapid Single Flux Quantum devices \cite{Michal2009} and sub-mm wave receivers \cite{Koshelets2000}, offer higher operating frequencies and less parasitic effects, especially if the object of interest can be integrated into the matching circuit on chip, as is possible with devices based, for example, on carbon nanotubes, semiconductor nanowires, or single molecules.
With this approach, parasitic effects can be minimized and compensated for in the design.
Operating frequencies in the 2-10 GHz range can be realized, allowing the use of measurement techniques developed in the context of circuit quantum electrodynamics \cite{Wallraff2004b} and for quantum limited amplifiers \cite{Bergeal2010,Castellanos2007}.

Following this paradigm, this paper presents distributed element impedance transformers into which a gold break junction has been integrated. Using the controlled electromigration \cite{Park99} of the break junction to realize a variable load, measured RF spectra of the device at different break junction resistances are presented and it is shown that a simple equivalent model yields quantitative agreement between the measured direct current (DC) resistance and RF impedance of the break junction.

\begin{figure}[h!]
	\centering
		\includegraphics[width=\columnwidth]{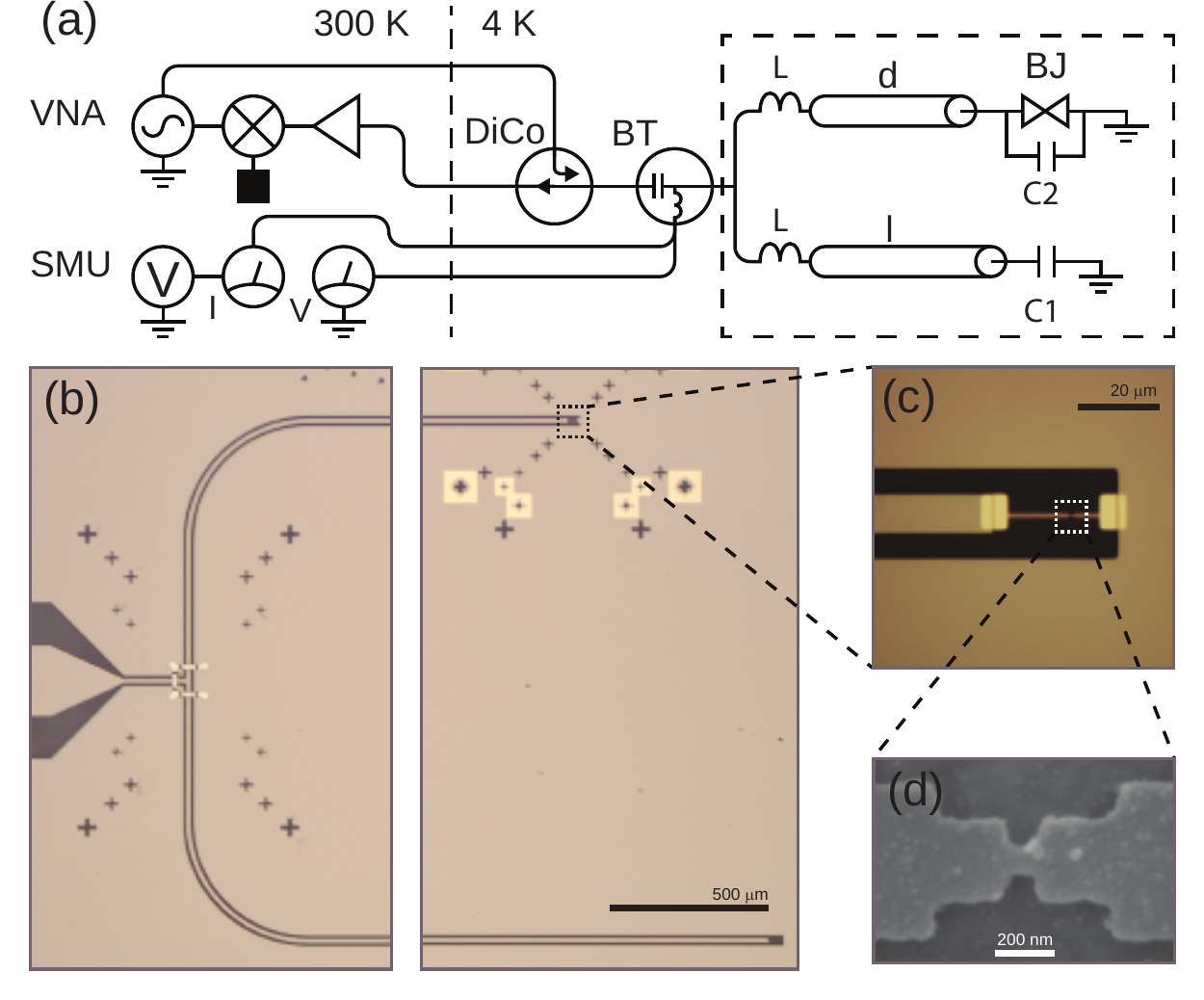}
	\caption{(a) Schematic of the circuit (dashed box) and setup: The break junction (BJ) is connected to a transmission line of length d, which is shunted by a line of length l, with parasitic inductances L from the T-junction and capacitances C1 and C2 at the ends of the transmission line.  A bias-tee allows DC biasing of the junction using a source-measure-unit (SMU). A directional coupler (DiCo) allows the measurement of the reflection coefficient using a vector-network analyzer (VNA). (b) Optical microscope image of the described device. The input port (left) splits into the upper transmission line to which the break junction (black rectangle) is connected and the open stub (lower line). Airbridges are used to connect the groundplanes at the T-junction. (c) Image of the break junction showing the end of the transmission line, the contact pads (gold rectangles), and the break junction (gold-red line). (d) SEM Image of the break junction. The wire is 70 nm wide.}
	\label{fig:chip}
\end{figure}

The presented resonant circuits consist of two lengths of transmission line connected in parallel, one terminated by the load, the other one by an open circuit (Fig. \ref{fig:chip} (a) dashed box), a configuration referred to as a single-stub shunt tuning circuit \cite{Pozar05,Hellmueller2012}, or stub tuner. This circuit makes use of the fact that a piece of transmission line transforms the impedance of an attached load, depending on the length of the line. Starting with a design load \Zd, also referred to as matched load, a length of transmission line \Ld\ (upper transmission line in Fig. \ref{fig:chip} (a)) is chosen such that the admittance of the combined line and load is of the form  $Y = 1/\Zn + iB$, with \Zn\ the characteristic impedance of the transmission line and $B$ a susceptance. By shunting the line at this length \Ld\ with an open transmission line of length \Ll\ (lower transmission line in Fig. \ref{fig:chip} (a)) and admittance $-iB$, the susceptances cancel leaving a combined impedance of \Zn, matching the feed line. If the load impedance is changed from the design load \Zd, the reflection coefficient \reflc\ does also change, allowing the load to be inferred from a measurement of the reflection coefficient.
In a lumped element approximation, the transmission line of length \Ld\ acts as a series capacitor, while the shunt line acts as a shunt inductor. To achieve this behavior, \Ld\ is chosen to be shorter than \qwl, while \Ll\ is chosen to be longer than \qwl, with both lengths approaching \qwl\ in the limit of large \Zd.

A simple model of the circuit impedance $Z_{\rm{Total}}$, from which \reflc\ can be calculated, is given by the parallel sum of the impedances presented by each line:
\begin{displaymath}
	Z_{\rm{Total}} = \Zn \Bigg(\tanh(\gamma \Ll)+ \frac{\Zn + \Zload \tanh(\gamma \Ld)}{\Zload + \Zn \tanh(\gamma \Ld)}\Bigg)^{-1}
\end{displaymath}
where $\gamma = \loss + i \pvelo$ is the propagation constant, with \loss\ representing the attenuation constant and \pvelo\ the phase constant of the transmission line. Using this model and numerical optimization, the necessary lengths for a chosen matched load can be obtained. Practically, the maximum matched load is limited by the attenuation constant of the transmission lines and the fabrication accuracy. While the losses extracted from coplanar resonator measurements \cite{Goppl08} allow matching in the G\Ohm\ range at 6 GHz and 100 mK, our current fabrication tolerances limit the realizable range to a \Zd\ of 1-10 M\Ohm. Care needs to be taken to include parasitic effects, both stray capacitances at the end of the lines, as well as stray inductances at the T-junction, as both of these effects modify the effective lengths of the transmission lines \cite{Simons2001}.

The presented devices are realized in 150 nm thick niobium coplanar waveguides on a 500 \micron\ thick sapphire substrate \cite{Simons2001}. Starting with a commercially coated 2\texttt{"} wafer, the transmission lines are patterned using photolithography and reactive ion etching (Fig. \ref{fig:chip} b). To ensure good electrical contact between the niobium film and the break junctions fabricated in a subsequent step, contact pads are first defined via photolithography, argon-ion etched to remove any oxide, and metalized with 5 nm titanium and 50 nm gold deposited using evaporation at an $8^\circ$ angle while rotating the substrate at $20^\circ$/s (bright gold pads in Fig. \ref{fig:chip} (c)). The  20 nm thick gold break junctions are patterned by electron beam lithography and lift-off and are 70-100 nm wide at their most narrow point (Fig. \ref{fig:chip} (d)). To suppress spurious modes of the coplanar transmission lines, 250 nm aluminum / 250 nm titanium bilayer airbridges are positioned at the T-junctions in a two-step photo-lithography and lift-off process. After dicing the wafer, the chips with four stub tuners each are cleaned with NMP/Acetone/IPA/O2 plasma, mounted in an RF printed circuit board and wire bonded.

The measurement setup consists of a hermetically sealed copper sample holder box, to which two copies of the same circuit are attached, a bias-tee for applying DC bias followed by a directional coupler to apply a microwave tone and measure the reflection coefficient (Fig. \ref{fig:chip} (a)). These components are mounted at the end of a dip-stick, allowing immersion in liquid helium, and are connected to a room-temperature feed-through via twisted pairs and semi-rigid coaxial cables. For DC measurements, a source-measure-unit is used in a 4 point configuration up to the bias-tee, yielding a systematic error of less than 1 \Ohm\ for resistance measurements. RF measurements are performed with a vector network analyzer (VNA), at a source power of -25 dBm and an IF bandwidth of 1 kHz. A short/open/load/thru calibration was performed for the flexible SMA coax lines connecting the VNA to the dipstick. Both instruments are controlled using custom software.

\begin{figure}
	\centering
		\includegraphics[width=1.00\columnwidth]{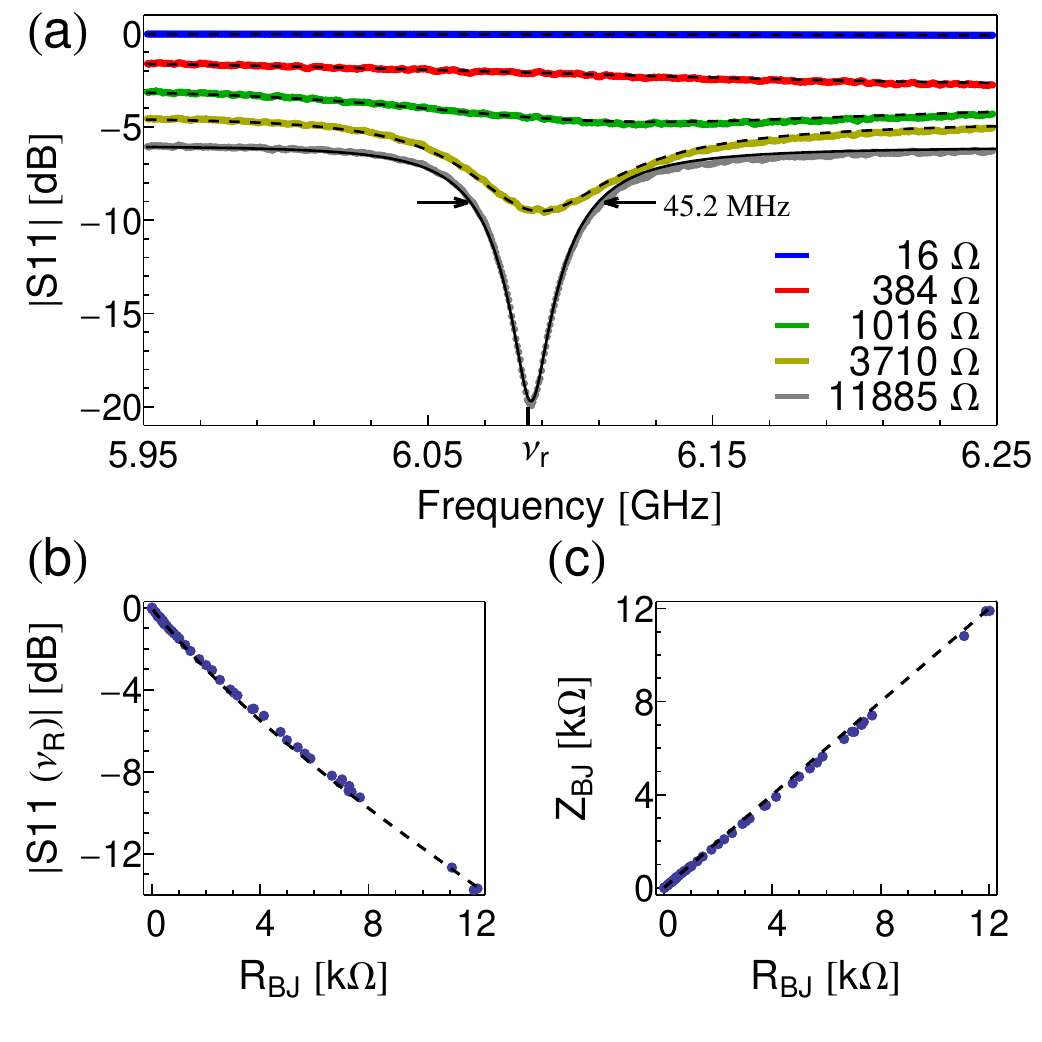}
	\caption{(a) Reflectance spectra \reflcabs\ taken at different resistances of the break junction, the spectra are offset by 1.5 dB each for better visibility. The fit extracting the electrical parameters of the circuit is shown by the solid line, the fits for determining the RF impedance of the break junction are shown as dashed lines. (b) The measured reflection coefficient \reflcabs\ (blue dots) plotted vs. the measured DC resistance at \fres\ = 6.085 GHz. The dotted line represents the reflection coefficient calculated using the extracted parameters. (c) Extracted RF impedance \zbj\, plotted vs. measured DC resistance \rbj , the dashed line indicates $\zbj = \rbj$ .}
	\label{fig:measurement}
\end{figure}

To characterize the devices with a variable load, the resistance of the integrated break junction is controlled by electromigration \cite{Park99}, and RF reflection spectra are taken for different resistance values. A simple feedback algorithm is used to control the electromigration by evaluating the change in resistance with respect to a reference resistance. This proceeds as follows: An initial reference resistance is measured at the start of the migration procedure, followed by ramping up the applied voltage at 5 mV/s. During this ramp, the resistance is monitored and compared to the reference resistance. Once a given threshold is reached, in the presented case a 5 \% change, the voltage is reduced by 30 mV, a new reference resistance is taken and the voltage ramp is applied again. This feedback method allows a controlled migration of the junction. At selected resistance values, the bias is reduced to 10 mV, a value at which effectively no migration occurs, an RF spectrum is taken and the algorithm restarted. This method allows controllable migration into the regime where quantized conductance is observed. It is even possible to migrate the junction to a single conductance quantum \Gn, a resistance of 12.9 k\Ohm, which is the signature of a single gold atom bridging the gap \cite{Scheer1997}. Further voltage ramps can be used to break the junction, yielding two electrodes separated by a nm-spaced gap with a tunneling resistance in the G\Ohm\ range.

To account for the microwave response of the setup itself and extract the reflection coefficient \reflc\ of the device only, the measured spectra are calibrated by subtracting a constant background. This background is obtained as the difference between the first measurement taken at a resistance of 16 \Ohm\ and the expected curve given by the model using the measured DC resistance.

Several designs with different matched loads were fabricated and measured and the results of a device with a design matched load of one \Gn\ are presented here. Furthermore, undercoupled $\lambda$/2 coplanar waveguide resonators were fabricated on the same wafer for an independent measurement of both attenuation and phase constant of the transmission lines.
A set of \reflcabs\ spectra for different measured DC resistances \rbj\ of the break junction are shown in Fig. \ref{fig:measurement} (a). At low \rbj, the circuit is out of resonance and fully reflects the incoming wave. As \rbj\ increases, the reflectivity decreases and starts to form a resonance, which becomes more pronounced as the load approaches the matched load. At a load resistance of of 12 k\Ohm, closest to the matched load, a bandwidth of 45.2 MHz is extracted.

The resonant circuit is characterized by the electrical lengths \Ld\ and \Ll\ as well as the loss factor \loss. The parasitic elements shown in Fig. \ref{fig:chip} (a) are modeled as variations in electrical length. As a reference, these three parameters are extracted from a fit to the measured spectrum for the largest insertion loss using the value of \rbj\ measured at DC (gray curve and solid line in Fig. \ref{fig:measurement} (a)). For all remaining measurements at other values of \rbj\ we determine the RF impedance \zbj\ of the break junction for each measured spectrum from a fit using the fixed set of parameters determined from the reference measurement (dashed lines in Fig. \ref{fig:measurement} (a)). This allows us to characterize how well the model describes the actual device characteristics.

Using the described calibration method and model, the electrical lengths, not taking into account parasitic effects,  were determined to be  \Ll\ = 5300.8 $\pm$ 0.2 \micron\ and \Ld\ = 4853.4 $\pm$ 0.2 \micron, close to the design values of \Ll\ = 5268.4 \micron\ and \Ld\ = 4852.4 \micron. The extracted attenuation constant is \loss\ = 0.050, comparable to the value extracted from the resonator measurement of \loss\ = 0.023. Part of the deviation from the expected values is due to the influence of the  parasitic elements mentioned before, while another part is due to impedance mismatches in the setup causing reflections, which are not taken into account in the calibration procedure. Using these extracted lengths, the resonance frequency \fres\ of the device was determined to be 6.085 GHz, with a bandwidth of 37.5 MHZ at a matched load of 29.3 k\Ohm. Both the parasitic inductance $L\approx$ 2 pH of the T-junction and the parasitic capacitance $C2\approx$ 1 fF of the break junction was approximately determined from measurements of a similar device with less bandwidth.

As shown in Fig. \ref{fig:measurement} (a), the fit of the model to the data shows good agreement, both for the fit extracting the characteristic parameters (solid line) as well as the fits from which the RF impedance \zbj\ of the break junction is extracted (dashed lines). The measured reflection coefficient $|S11|$\ at \fres\ also shows good agreement compared with the calculated values using the model and extracted lengths, as shown in Fig. \ref{fig:measurement} (b).
To further characterize the properties of the device and its description by the model, we plot the fitted RF impedance \zbj\ vs. the measured DC resistance \rbj\ in Fig. \ref{fig:measurement} (c). This also yields good quantitative agreement, showing the expected equivalence between \rbj\ and \zbj\, measured at DC and extracted from RF reflectometry. The small deviations in the 4-8 k\Ohm\ range can be attributed to the inaccuracies in the calibration.

An upper bound for the attainable time resolution of a reflectometry measurement is set by the maximum band width of the employed impedance matching circuit. To which degree this upper bound can be approached, depends on the nature of the signal to be investigated and the signal to noise ratio achievable with a given signal strength and amplification chain. If the signal of interest can be reproduced a large number times, such as in many circuit QED experiments, e.g.~Ref.~\cite{Bozyigit2011}, the full bandwidth of the circuit can be exploited by repeating the measurement to achieve the wanted signal to noise ratio. For random signals, the time resolution is typically limited by the noise added by the amplification chain, such that the full bandwidth of the circuit may not be reached at the desired signal to noise ratio, as for example in electron counting experiments \cite{Gustavsson2006}. Using quantum limited parametric amplifiers \cite{Bergeal2010,Castellanos2007} may enable approaching the maximum temporal resolution possible with a given matching circuit.

In conclusion we have realized a superconducting impedance matching circuit implemented with distributed elements and characterized it using an integrated gold break junction. The measured reflectance spectra for different resistances of the junction are described by an equivalent circuit with good agreement between model and measurement. Furthermore we have extracted the real part of the impedance of the break junction from an RF measurement and find good agreement with the measured DC resistance. The device demonstrated here may be used for time-resolved reflectometry measurements performed on nano-scale objects such as break junctions or single molecule devices.

The authors would like to thank Peter Leek, Andreas Alt and the Staff at FIRST for technical support.

\end{document}